\documentstyle[12pt,aaspp4]{article}

\def\kms{km~s$^{-1}$}
\def\hmpc{$h^{-1}\,$Mpc}
\def\hkpc{$h^{-1}\,$kpc}
\def\etal{{et~al.}} 
\def\sn{\ifmmode{S_N}\else$S_N$\fi}
\def\pf{\ifmmode{{\hbox{\sc psf}}}\else{{\sc psf}}\fi}

\begin{document}

\title{The Dependence of Globular Cluster Number on Density
for Abell Cluster Central Galaxies\altaffilmark{1}}

\author{John P. Blakeslee\altaffilmark{2}}
\affil{Department of Physics, 
	Massachusetts Institute of Technology,
	Cambridge, MA 02139}
\authoremail{jpb@astro.caltech.edu}

\altaffiltext{1}{Observations conducted at the
Michigan-Dartmouth-MIT Observatory}
\altaffiltext{2}{Current address:  Palomar Observatory, California Institute
of Technology, Mail Stop 105-24, Pasadena, CA 91125; jpb@astro.caltech.edu}
 
\begin{abstract}
A study of the globular cluster systems of 23
brightest, or central, galaxies in 19 Abell clusters has
recently been completed (Blakeslee 1997).
This letter presents some of the newly discovered correlations of
the globular cluster specific frequency $S_N$ in these galaxies
with other galaxy and cluster properties and puts forth an
interpretation of these results.
$S_N$ is found to correlate strongly with measures of the cluster
density, such as the velocity dispersion
of the cluster galaxies and the cluster X-ray temperature and luminosity
(especially ``local'' X-ray luminosity).
Within a cluster, 
galaxies at smaller projected distances from the X-ray center
are found to have higher values of $S_N$.
Taken together, the scaling of $S_N$ with cluster density and
the relative constancy of central galaxy luminosity
suggest a scenario in which globular clusters form in proportionate numbers
to the available mass, but galaxy luminosity ``saturates'' at a 
maximum threshold, resulting in higher \sn\ values for central galaxies
in denser clusters as well as the suitability of these galaxies as
``standard candles.''
Thus, these galaxies do not have too many globular clusters for
their luminosity, they are underluminous for their number of 
globular clusters.
\end{abstract}


\clearpage

\section{Introduction}

The number of globular clusters (GCs) per unit galaxy luminosity is known
as the globular cluster ``specific frequency''
\sn\ (\cite{hv81}).
Some central dominant galaxies in clusters, such as M87 in Virgo, have
huge GC populations, or systems (GCSs), even when their large luminosities
are taken into consideration (see \cite{h91} and references therein).
These galaxies are often called ``high-\sn\ systems'' and 
sometimes described as having ``excess'' GC populations.
Although the high-\sn\ systems are usually viewed as a special
class of object, whether they represent the upper end of a continuum or are
truly distinct has not previously been established. 
Because of the rarity of these objects, there have been few observational
constraints on theories of their formation.

In an effort to address this situation,
Blakeslee \& Tonry (1995) developed an analysis technique
for studying GCSs at relatively large distances and used it
to study the GCSs of the two giant central galaxies in the Coma cluster.
The technique is adapted from the methods used in the surface brightness
fluctuations distance survey (\cite{sbf97}), with
the basic idea being that the portion of the GCS below the limit of
direct photometric detection still produces a measurable signal in
the power spectrum of the galaxy image.
Wing \etal\ (1995) independently
used the same approach to study the M87 GCS.

We have applied this technique to study a large, complete
sample of brightest galaxies in Abell clusters.  
Our intent was to learn which central galaxies are high-\sn\ systems,
which are not, and why.  
A complete description of all aspects of this project is
given by Blakeslee (1997; hereafter B97).
Most of the details and the results
will be published in a forthcoming paper 
(Blakeslee, Tonry, \& Metzger 1997; hereafter \cite{btm}). 
The present Letter summarizes the observational sample,
presents some of the most provocative new correlations
of central galaxy \sn\ with properties of the galaxy clusters,
and offers a new interpretation of the high \sn\ galaxies.
Unlike other models which have attempted to explain \sn\ variations
in terms of GC formation efficiency, this interpretation sees
the relative insensitivity of galaxy luminosity to environmental density
as the main culprit behind the large variations in \sn.

\section{The Data}

The details of the sample selection, observations, and reductions
are given elsewhere (\cite{thesis}) and will
be discussed again by \cite{btm}.
Here we only summarize these aspects of the project
and concentrate on the trends of
\sn\ with cluster properties for our sample of Abell cluster
central galaxies.  

The sample of galaxies studied here was selected from the 
Lauer \& Postman (1994; hereafter LP) survey of 119 brightest cluster
galaxies (BCGs) in Abell clusters and supplemented by the
partial list of second brightest, or ranked, cluster galaxies (SRGs)
from Postman \& Lauer (1995; hereafter PL).
Specifically, the current sample includes all early-type
BCGs in northern Abell clusters with $cz < 10000$~\kms, 
absolute $R$-band metric magnitude
$M_R < -22.0 + 5\log h_{75}$, and galactic latitude $|b| > 15^{\circ}$.
(The early-type and $M_R$ criteria exclude just two Abell clusters each.)
It also includes three second brightest galaxies, and one third brightest,
because these galaxies were similar to the respective BCGs in
luminosity and extent.  The final sample then comprises 23 galaxies in
19 Abell clusters, including the two Coma galaxies studied by
Blakeslee \& Tonry (1995).
%
The observations were conducted with the 2.4~m telescope at the
Michigan-Dartmouth-MIT Observatory on Kitt Peak over the course
of several observing runs.  
Total integration times on the sample galaxies ranged from just over 1~hr
to nearly 6.5~hr, determined by the prevailing image quality, the galaxy
distance, and the available time.  

The reduction process for the images is basically the same as that described
by \cite{bt95}.  For each final galaxy image, we model and subtract the galaxy
light, then run the automatic photometry program DoPhot (\cite{dpo}).
After correcting for the presence of dwarf and background galaxies in the
object counts, we arrive at an estimate of the total number of GCs brighter
than the cutoff magnitude $m_c$ (determined by artificial star tests).
We then remove all objects brighter than $m_c$ from the
image and measure the power spectrum of what remains.
In order to estimate the total number of GCs fainter than $m_c$,
the power spectrum measurement must be corrected for
background galaxies and stellar SBF; the contribution from possible
faint dwarf galaxies in the halo of the BCG is negligible at these magnitudes
(see \cite{btm}).

The DoPhot object counts and power spectrum measurement
each provide an estimate of the
total number of GCs, given a luminosity function (GCLF); dividing the total
number of GCs by the galaxy luminosity yields \sn.
We assume the M87 GCLF (\cite{m87hst})
and calculate a weighted average of the two values of 
\sn\ derived for each galaxy, one from the
counts and the other from the power spectrum.  The agreement between
the two separate measurements indicates that the use of the M87 GCLF is
justified (\cite{thesis}; \cite{btm}).

\section{Specific Frequencies}

\subsection{Results for Individual Galaxies}

Table~\ref{tab:sn} lists the values of \sn\ for each of the program
galaxies, identified by the Abell cluster number and the luminosity rank in
the cluster.  In order to avoid uncertain extrapolations and redshift biases
in \sn, the table gives ``metric \sn'' values, calculated within the same
physical radius of 32~\hkpc\ of the center of each galaxy, excluding the
unusable galaxy center.  The metric radius of 32~\hkpc\ was chosen because
it corresponds to the limit of the image for the nearest of the sample
galaxies (roughly 500~pix, or 2\farcm3, in the 1024$^2$ images).  In
calculating \sn, we adopted a zero point for the galaxy luminosities based
on a Virgo distance of 16~Mpc.  The distance zero point is unimportant,
however, as any observed trends will be independent of it.

The tabulated \sn\ uncertainties include contributions from measurement
error, relative distance error due to an assumed one-dimensional rms
peculiar velocity of 400~\kms\ for the galaxy clusters in the CMB frame, and
intrinsic dispersions of $\pm$0.20 and $\pm$0.05~mag for the mean and width
of the GCLF, respectively (\cite{h96}; \cite{thesis}).  Uncertainties due to the
distance, velocity, or GCLF calibrations, which could systematically change
all the values by perhaps 20\%, are not included.  The \sn\ values shown for
A1656-1 (NGC~4889) and A1656-2 (NGC~4874) differ from the global values
reported by Blakeslee \& Tonry (1995) primarily due to the faulty RC2
photometry used in that study, but also because the earlier values were
calculated within a larger radius.

As evident from the table, 
there is no separation of ``normal'' and ``anomalous'' \sn\ classes
in this sample of Abell cluster central galaxies. 
\sn\ varies uniformly from $\sim\,$3 to $\sim\,$9. 
In the three clusters represented by more
than one member, the SRGs all
have the higher \sn\ values, though the difference is not significant
in the case of A539.  This is a result of selection.  The SRGs were
included in this sample because they all 
appear to dominate their clusters at least as much as the actual
BCGs selected by LP.  Moreover, these galaxies all have more extended
profiles and are closer to the cluster X-ray centers
than their respective BCGs.
The relevance of these points will be clarified below.

\subsection{\sn\ and Cluster Density}

\begin{figure}\epsscale{0.7}
\plotone{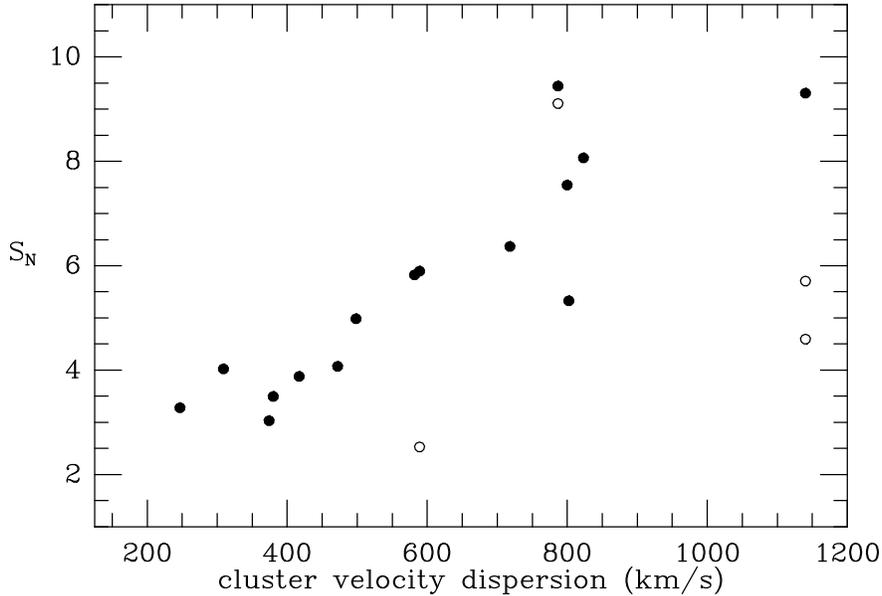}\caption{\small
The derived $S_N$ values are shown plotted against cluster velocity
dispersion, collected from the literature,
within one Abell radius of the cluster center.
Filled circles represent the central bright galaxies in the clusters;
open circles represent less central galaxies for clusters with more than
one member in the sample (see text for details).
\label{fig1}}
\end{figure}

A major result of this project is the strong dependence
found for central galaxy \sn\ on cluster density.
Figure~\ref{fig1} shows the \sn\ values from Table~\ref{tab:sn}
plotted against the Abell cluster velocity dispersions,
from Beers \etal\ (1991), Girardi \etal\ (1993), Scodeggio \etal\ (1995),
Struble \& Rood (1991), and Zabludoff \etal\ (1993).
For the three clusters with multiple members,
the more central galaxies (judged by the cluster X-ray center)
are the ones shown as the filled symbols, and the less central ones
are shown as open circles, even though three of these were 
chosen as BCGs by \cite{lp}.  The figure clearly
indicates that central galaxies in higher dispersion clusters,
and thus at the centers of deeper potential wells,
have significantly more GCs per unit luminosity than those in
lower dispersion clusters.  
(The significance of the correlation is near 1.0 for the filled symbols.)

\begin{figure}\epsscale{0.7}
\plotone{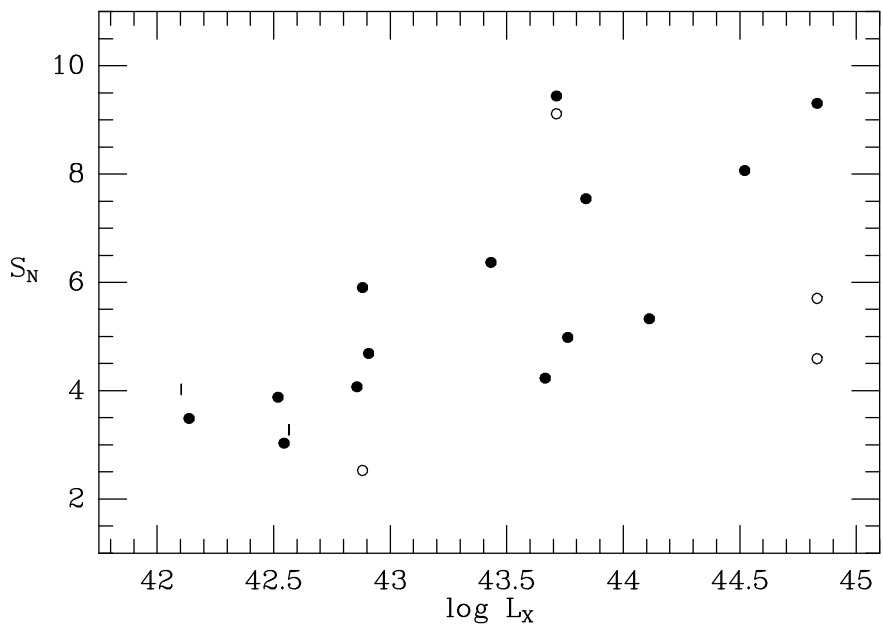}\caption{\small
$S_N$ is plotted against the total X-ray luminosity
(ergs~sec$^{-1}$) within 0.5~\hmpc\ of the cluster X-ray center.
The two short vertical lines represent upper limits on the
X-ray luminosity; otherwise, symbols are as in Figure~\ref{fig1}.
\label{fig2}}
\end{figure}

Perhaps the best measure of the depth of the cluster potential is
the temperature $T_X$ of the intracluster X-ray emitting gas.
However, relatively few of these clusters have had this quantity
directly measured, so we use instead the cluster X-ray luminosity $L_X$.
Figure~\ref{fig2} plots \sn\ against the logarithm of $L_X$ measured
in the 0.5--4.5~keV band from Jones \& Forman (1997).
While previous investigations found no correlations
of \sn\ with X-ray properties (Harris \etal\ 1995; West \etal\ 1995),
our larger, more homogeneous data set clearly shows that there is
a correlation.

Recently, West \etal\ (1995) have discussed the possibility of
a population of ``intra\-cluster globular clusters'' (IGCs) which
belong to a galaxy cluster as a whole, not to individual galaxies.
The hypothetical IGCs trace the cluster mass profile; 
assuming hydrostatic equilibrium and an isothermal potential implies
that their surface density will be proportional to
$T_X/(1{\,+\,}r^2/r^2_c)$, where $r$ is 
the projected distance of the galaxy from the cluster X-ray center,
and $r_c$ is the core radius.
In this scenario, a large galaxy near the center of a cluster
will appear to have an elevated value of \sn\ because some of these IGCs
will have become associated with it.

\begin{figure}\epsscale{0.7}
\plotone{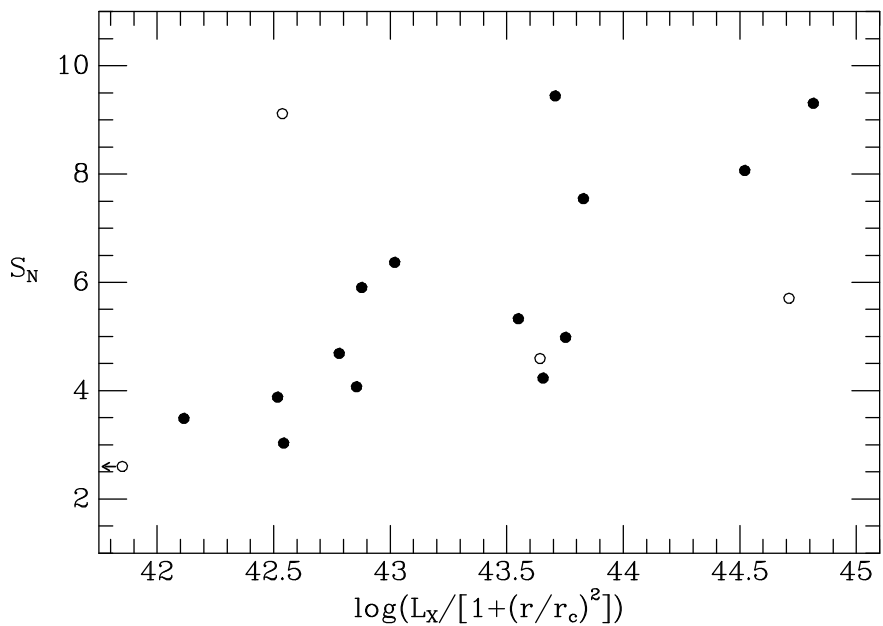}\caption{\small
$S_N$ is plotted against the radially weighted, or ``local'',
X-ray luminosity at the galaxy position within the cluster. 
The correlation observed in Figure~\ref{fig3} improves slightly
with this weighting.  The point for A2197-1, shown with an arrow,
would be located at log$[L_X/(1{\,+\,}r^2/r^2_c)] = 40.8$.
\label{fig3}}
\end{figure}

Guided by West \etal, we plot $S_N$ against
$L_X/(1{\,+\,}r^2/r^2_c)$, the ``local X-ray luminosity'',
in Figure~\ref{fig3}. 
(This differs from the corresponding figure in
\cite{thesis} because a calculational error in that work 
resulted in the use of core radii that were systematically too large.)
The scatter in the relation
decreases by $\sim\,$5\% with this radial weighting.
However, for reasons discussed by \cite{thesis}, i.e.,
the theoretical problems in producing the IGC population
and the relatively low velocity dispersion observed in the
very high \sn\ GCS of M87, we prefer to think of all the GCs as
being bound to the host galaxy.  The improvement of the 
\sn--$L_X$ correlation with the radial weighting would then 
be related to environmental effects having to do with the
relative dominance of the galaxy in the cluster potential.

The correlations shown here between the \sn\ value of the
central bright galaxy in the cluster and measures of the depth
of the cluster potential might be explained in a number of ways.
For instance: (1) the efficiency of GC formation for galaxies in
denser regions may have been higher, for whatever reason; (2) central
galaxies in clusters may strip GCs over time from other cluster members;
or (3) the number of GCs scales with the cluster mass, but central 
galaxy luminosity does not, 
resulting in higher values of \sn\ {\it and} the observed
independence of BCG luminosity on cluster properties. 
This final possibility has the aesthetic advantage of combining two
separate problems into one.
This issue is discussed more in the following section.

\section{Discussion}

We have established that central cluster galaxy \sn\ scales with
measures of Abell cluster size, in particular, velocity dispersion
and X-ray properties. (B97 shows that it also correlates to a lesser
degree with cluster richness, but not morphology.)
We have also found that for clusters with multiple galaxies in
this sample, the one with the higher \sn\ value
is  the one which was closer to the cluster X-ray center.

\begin{figure}\epsscale{0.75}
\plotone{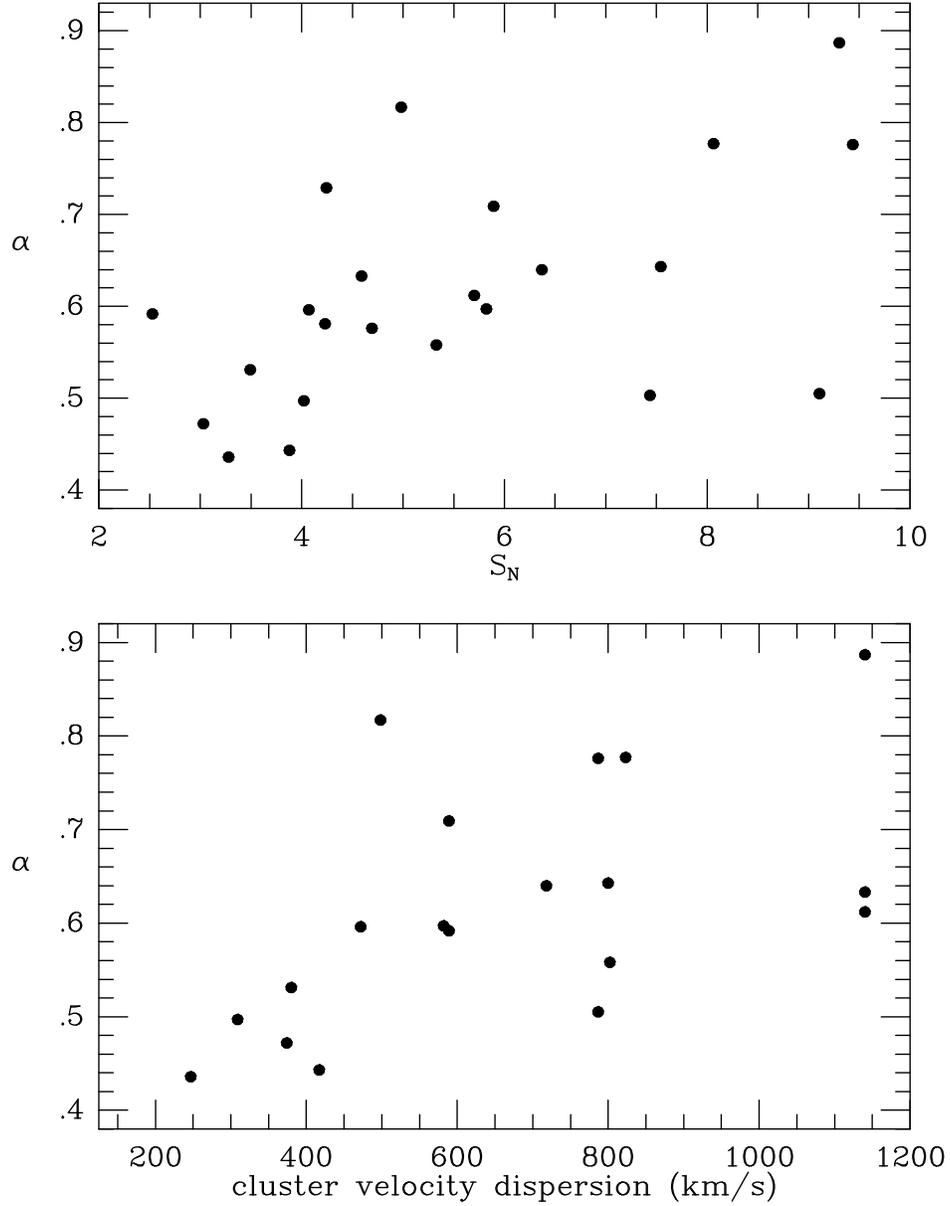}\caption{\small
The profile structure parameter $\alpha$, a measure of galaxy extent,
is plotted against \sn\ (top) and cluster velocity dispersion (bottom).
The correlation between \sn\ and $\alpha$ appears to be a consequence
of the fact that both are dependent on cluster density, reflected by the
velocity dispersion.
\label{fig4}}
\end{figure}

So far, we have not discussed correlations of \sn\ with properties
of the host galaxy.  The obvious property to choose is galaxy luminosity,
but due to the relatively small variation in this quantity among BCGs
(\cite{pl} and references therein), any correlation of \sn\ with luminosity
for this sample of galaxies is marginal
(\cite{thesis}; see also \cite{h91}).   A better, though still
comparatively weak, correlation is found with galaxy extent,
as measured, for example, by the profile
``structure parameter'' $\alpha$, the logarithmic slope of the
integrated galaxy luminosity evaluated at $r{\,=\,}10$ \hkpc.
It should be recognized, however, that $\alpha$ was first introduced
by Hoessel (1980) in order to remove the weak richness dependence
of the BCG ``standard candle'' distance indicator.  Figure~\ref{fig4}
illustrates this situation, with $\alpha$ plotted against \sn\
in the top panel and against cluster velocity dispersion
in the lower panel.  The two correlations are of similar significance
levels; both have much greater scatter than the relation between
\sn\ and cluster density or X-ray luminosity.

B97 has advanced a unified view of the observed
\sn\ variations and the ``standard candle'' aspect of BCGs.
In this view, GCs form early and in proportion to
the local mass density, thus the observed excellent correlations of their
number with cluster velocity dispersion and X-ray properties.
(There is no need to invoke dark matter biasing.)
On the other hand, the luminosity of the central galaxy saturates,
speculatively  due to the process of cluster formation itself.
The collapse and subsequent
virialization of a cluster from many smaller, irregular groups of
galaxies might well disturb the gas in a large, centrally
located galaxy in such a way as to halt subsequent star formation, adding
to the hot intra\-cluster gas and perhaps leaving as its
signature the characteristic shallow profiles of these galaxies
(i.e., the observed correlation with $\alpha$).
 
Although the scenario remains uncertain, it is clear 
that central galaxy GC number depends
on cluster scale, and can therefore itself be considered
a cluster property, while the galaxy luminosity does not. 
Our proposal is that the same situation gives 
rise to both the high values of \sn\
measured for some galaxies, and the suitability of BCGs as ``standard
candles.''   In light of these results, we 
suggest that the proper galaxy to use for measuring distances
based on the $L_m$-$\alpha$ indicator
is the bright extended galaxy closest to the cluster center;
we predict that this galaxy will also be the one with the highest \sn.

\acknowledgments

I thank my thesis advisor John Tonry for insight-filled discussions.
This research was supported by NSF grant AST94-01519 and by a
Caltech Fairchild Fellowship.

\begin{deluxetable}{ll|ll|ll}
\small
\centering
\tablewidth{0pt}
\tabcolsep=0.3cm
\newdimen\digitwidth
\setbox0=\hbox{\rm0}
\digitwidth=\wd0
\catcode`?=\active
\def?{\kern\digitwidth}
\tablecaption{Metric Specific Frequencies for Sample Galaxies\label{tab:sn}}
\vspace{-2truemm}
\tablehead{\colhead{Galaxy} & \colhead{$S_N\;\,^+_-$} &
\colhead{Galaxy} & \colhead{$S_N\;\,^+_-$} &
\colhead{Galaxy} & \colhead{$S_N\;\,^+_-$} }
\startdata
?A262-1 & $5.0\;^{1.5}_{1.3}$ & ?A999-1 & $3.9\;^{1.5}_{1.3}$ & A1656-3 & $4.6\;^{1.5}_{1.3}$ \\
?A347-1 & $5.8\;^{1.6}_{1.3}$ & A1016-1 & $3.3\;^{1.2}_{1.1}$ & A2162-1 & $7.4\;^{2.2}_{1.8}$ \\
?A397-1 & $4.7\;^{1.4}_{1.1}$ & A1177-1 & $4.2\;^{1.3}_{1.0}$ & A2197-1 & $2.5\;^{1.4}_{1.3}$ \\
?A539-1 & $9.1\;^{3.0}_{2.6}$ & A1185-1 & $6.4\;^{1.8}_{1.4}$ & A2197-2 & $5.9\;^{1.5}_{1.2}$ \\
?A539-2 & $9.4\;^{3.0}_{2.4}$ & A1314-1 & $4.2\;^{1.1}_{1.0}$ & A2199-1 & $8.1\;^{2.3}_{1.9}$ \\
?A569-1 & $3.0\;^{1.2}_{1.0}$ & A1367-1 & $5.3\;^{1.4}_{1.1}$ & A2634-1 & $7.5\;^{2.1}_{1.7}$ \\
?A634-1 & $4.0\;^{1.2}_{1.0}$ & A1656-1 & $5.7\;^{1.3}_{1.3}$ & A2666-1 & $3.5\;^{1.1}_{1.0}$ \\
?A779-1 & $4.1\;^{1.0}_{0.9}$ & A1656-2 & $9.3\;^{2.0}_{2.0}$ &   &     \\
\enddata
\vspace{-0.6truecm}
\tablecomments{
\sn\ values are calculated within 32~\hkpc\ of the center of
each galaxy assuming the M87 GCLF shifted according to the cluster CMB frame
velocity; a CMB velocity of 1310~\kms\ has been adopted for Virgo.}
\end{deluxetable}
\vspace{0.4truecm}

\clearpage

\end{document}